\documentclass[useAMS,usenatbib]{mn2e}

\usepackage{amsfonts}
\usepackage{graphicx}
\usepackage{aas_macros} 
\renewcommand{\vec}[1]{\bmath{#1}}
\newcommand{\field}[1]{\mathbb{#1}}

\newcommand{\pder}[2]{\ensuremath{\frac{\partial{#1}}{\partial{#2}}}}
\newcommand{\pdder}[2]{\ensuremath{\frac{\partial^2{#1}}{\partial{#2}^2}}}
\newcommand{\pderc}[3]{\ensuremath{\left.\frac{\partial{#1}}{\partial{#2}}\right|_{#3}}}

\title[Weakly nonlinear spiral density waves]
{A Weakly nonlinear theory for spiral density waves excited by accretion disc
turbulence}
\author[T.~Heinemann and J.~C.~B.~Papaloizou]{
T.~Heinemann$^{1}$
\thanks{tobi@ias.edu}
and J.~C.~B.~Papaloizou$^{2}$
\thanks{J.C.B.Papaloizou@damtp.cam.ac.uk}\\
$^{1}$Institute for Advanced Study,
Einstein Drive,
Princeton, NJ 08540,
USA\\
$^{2}$Department of Applied Mathematics and Theoretical Physics,
University of Cambridge,\\
Centre for Mathematical Sciences, 
Wilberforce Road, Cambridge, CB3 0WA, UK }

\begin{document}

\date{Accepted Received ; in original form }

\pagerange{\pageref{firstpage}--\pageref{lastpage}} \pubyear{2011}

\maketitle

\label{firstpage}

\begin{abstract}
We develop an analytic theory to describe spiral density waves propagating in
a shearing disc in the weakly nonlinear regime. Such waves are generically
found to be excited in simulations of turbulent accretion disks, in particular
if said turbulence arises from the magneto-rotational instability (MRI). We
derive a modified Burgers equation governing their dynamics, which includes
the effects of nonlinear steepening, dispersion, and a bulk viscosity to
support shocks. We solve this equation approximately to obtain nonlinear
sawtooth solutions that are asymptotically valid at late times.
In this limit, the presence of shocks is found to cause the wave amplitude to
decrease with time as $t^{-2}$. The validity of the analytic description is
confirmed by direct numerical solution of the full nonlinear equations of
motion. The asymptotic forms of the wave profiles of the state variables are
also found to occur in MRI simulations indicating that dissipation due to
shocks plays a significant role apart from any effects arising from direct
coupling to the turbulence.
\end{abstract}


\section{Introduction}

Accretion discs are common in astrophysics, occurring in close binary systems,
active galactic nuclei and around protostars where they provide the
environment out of which planets form \citep[see e.g.][for
reviews]{PapaloizouLin1995,LinPapaloizou1996}.

Observational inferred accretion rates require enhanced angular momentum
transport to take place. This is believed to be mediated by some form of
turbulence, which can be regarded as providing an effective anomalous
viscosity. The most likely source of accretion disk turbulence is the
magneto-rotational instability \citep[MRI,
see][]{BalbusHawley1991,BalbusHawley1998}.

In both local and global simulations of the MRI with and without net magnetic
flux, prolific density wave excitation has been found (e.g.\
\citet{GardinerStone2005} in the local case, and \citet{Armitage1998} in the
global case). Such waves may be important in the contexts of quasi periodic
oscillations. They may also be associated with stochastic migration of
protoplanets \citep{NelsonPapaloizou2004,Nelson2005} and with providing some
residual angular momentum transport in magnetically inactive dead zones
\citet{Gammie1996}.

In view of the significance of these waves it is important to gain an
understanding of the processes leading to their excitation and ultimate
dissipation. In \citet{HeinemannPapaloizou2009a} we developed a WKBJ theory
for the excitation of these density waves in the linear regime and found that
vortensity fluctuations are responsible for their excitation during a short
period of time, as they change from being leading to trailing, denoted as
swing. In a second paper \citep{HeinemannPapaloizou2009b}, we studied the wave
excitation process directly as it occurs in fully non-linear three dimensional
numerical MRI simulations. The results were found to be in good agreement with
the WKBJ description of the excitation process developed in
\citet{HeinemannPapaloizou2009a}.

In this paper we extend the work of our previous two papers to consider the
behavior of the waves as they enter the nonlinear regime. This is significant
because the eventual development of weak shocks is expected and this leads to
the dissipation of the waves. Their maximum amplitude can be expected to be
determined by a balance between excitation and dissipation processes. The
latter might occur either directly through wave phenomena such as shocks, or
through interaction with the turbulence. Possibilities here include energy
loses resulting from random secondary density wave generation through
interaction of the primary wave with turbulent eddies
\citep[e.g.][]{Lighthill1953,Howe1971b,Howe1971a,FfowcsWillamsHowe1973}.
Another phenomenon that could potentially play a role is distortion of the
primary wave front on account of a variable wave propagation speed caused by
the presence of dynamic turbulent eddies
\citep[e.g.][]{HesselinkSturtevant1988}. Understanding these phenomena is
notoriously difficult and their theoretical description is, at present,
somewhat imprecise and associated with issues of interpretation. In the body
of this paper we focus on weakly nonlinear wave theory without turbulence and
defer consideration of wave-turbulence interactions until the discussion
section.

Our studies, in common with the majority of those that incorporate turbulence
resulting from the MRI, are restricted to local isothermal disc models.
Accordingly they do not incorporate the possibility of the channeling of wave
energy towards the upper disc layers, that may occur for density waves with
short wavelengths in the direction of the shear, in models with significant
thermal stratification \citep{OgilvieLubow1999}.

As shown in \citet{HeinemannPapaloizou2009a,HeinemannPapaloizou2009b},
turbulent vortensity fluctuations are able to excite pairs of
counter-propagating spiral density waves when they swing from leading to
trailing. Subsequently, as the excited waves propagate, their wavelengths in
the direction of propagation decrease and the conservation of wave momentum
results in an increasing amplitude, causing them to enter the nonlinear regime
in which the formation of shocks occurs. Our aim is to describe the time
asymptotic form of these excited waves and the way in which the associated
shock dissipation leads to amplitude decay, and to relate these phenomena to
what is seen in MRI simulations.

We consider the nonlinear evolution of a single pair of counter-propagating
spiral density waves on a homogeneous background. As we will demonstrate, this
becomes an effectively one-dimensional problem when described in a shearing
coordinate frame, reducing to ordinary one-dimensional gas dynamics in the far
trailing regime.

The plan of the paper is as follows. In Section~\ref{sec:model} we describe
the shearing box model and present the basic equations. In
Section~\ref{sec:shearcoord} we introduce shearing coordinates, giving the
form of the basic equations governing solutions that are functions only of the
shearing coordinate in the direction of the shear and time. We go on to
express these using a Lagrangian formalism in Section~\ref{sec:lagrange},
deriving the appropriate forms of specific vorticity or vortensity
conservation in Section~\ref{sec:vortensity}, and nonlinear wave momentum
conservation in Section~\ref{sec:wave-momentum}.

In Section~\ref{sec:analytic} we use the Lagrangian formulation as the basis
for the development of an analytic description that is applicable to the
weakly nonlinear regime. In this formulation, third order effects such as
entropy production in shocks are neglected, making it possible to adopt an
adiabatic or isothermal equation of state \citep[e.g.][]{Yano1996}. In
Section~\ref{sec:wave-equation} we derive a modified Burgers equation
governing the unidirectional propagation of weakly nonlinear waves \citep[for
a similar approach to the problem of nonlinear planetary wakes
see][]{GoodmanRafikov2001}. This is found to provide the correct nonlinear
development of wave profiles in which weak shocks are ultimately present.
Nonlinear sawtooth solutions, which are asymptotically valid at late times,
are derived in Section~\ref{sec:sawtooth}. Corrections to the wave profile
arising from dispersive effects are derived in Section~\ref{sec:dispersion}.

In Section~\ref{sec:simulations} we go on to present numerical solutions of
the full nonlinear equations governing a pair of counter-propagating spiral
density waves undergoing swing excitation. At late times these are compared
with analytic solutions derived in Section~\ref{sec:sawtooth}. The form of the
wave profiles and the decay of the wave amplitudes with time are found to be
in excellent agreement. Finally in Section~\ref{sec:discussion} we discuss our
results, comparing them with what is seen in, and considering their
consequences for, MRI simulations.

\section{Model basic equations and preliminaries}\label{sec:model}

We adopt the standard local shearing box of \citet{GoldreichLyndenBell1965}
and assume that the effect of magnetic fields on the density waves we consider
is negligible. The basic equations are the conservation of mass and momentum
in the form 
\begin{equation}
  \label{eq:continuity}
  \pder{\rho}{t} + \nabla\cdot(\rho\vec{v}) = 0
\end{equation}
\begin{equation}
  \label{eq:motion}
  \pder{\vec{v}}{t} + \vec{v}\cdot\nabla\vec{v}
  + 2\vec{\Omega}\times\vec{v} =
  - \frac{\nabla P}{\rho} + 2q\Omega^2 x\vec{e}_x + \vec{f}.
\end{equation}
Here $\rho$, $P$, $\vec{v}$, and $\vec{f}$ are the density, pressure,
velocity, and viscous force per unit mass, respectively. We adopt a Cartesian
coordinate system $(x,y,z)$ with origin at the center of the box. The
$x$-direction points away from a putative central object, the $y$-direction is
in the direction of rotation and the $z$-direction is parallel to the rotation
axis. The unit vectors in these coordinate directions are $\vec{e}_x$,
$\vec{e}_y$, and $\vec{e}_z$ respectively. The angular velocity of the system
is \mbox{$\vec{\Omega}=\Omega\vec{e}_z$}. For a Keplerian disk the constant
$q=3/2$.

In order to complete the description, we require an equation of state. This
may be found by noting that the pressure $P$ can be taken to be a function of
the density $\rho$ and the entropy per unit mass $S$. From the first and
second laws of thermodynamics, the entropy satisfies the equation
\begin{equation}
  T\left(\pder{S}{t} + \vec{v} \cdot \nabla S\right) = \epsilon,
\end{equation}
where $T$ is the temperature and $\epsilon$ is the net rate of heating per
unit mass. When the latter is zero -- as we will assume throughout this paper
-- then the motion is adiabatic, with $S$ being conserved on a fluid element,
and we may write $S=S_0$, where $S_0$ is the fixed entropy per unit mass. We
point out that heating by shocks will, in general, cause departures from an
adiabatic equation of state, even in the absence of cooling. However, such
effects enter only at third order in the wave amplitude
\citep[e.g.][]{Yano1996} and hence do not affect the dynamics of weakly
nonlinear waves of the type considered in this paper.

The basic state on which the waves we eventually consider propagate is such
that $P$, $\rho$ and $S$ are constant and the velocity
\mbox{$\vec{v}=-q\Omega{}x\vec{e}_y$}. Then for adiabatic motion $S_0$ is
constant and the relation $P=P(\rho,S_0)$ leads to an effectively barotropic
equation of state. We neglect vertical gravity and thermal stratification and
look for waves that have no dependence on $z$. We note, however, that in the
strictly isothermal case (\mbox{$P\propto\rho$}), the velocity amplitudes
associated with the density waves we consider turn out not to depend on $z$
even if the disk is vertically structured \citep{FromangPapaloizou2007}. But
note that this situation could be modified for density waves with short
wavelength in the direction of the shear if the disc model were to have
significant thermal stratification \citep{OgilvieLubow1999}.

\subsection{Shearing coordinates}\label{sec:shearcoord}

In the theory of linear wave excitation developed in
\citet{HeinemannPapaloizou2009a,HeinemannPapaloizou2009b} the coordinate
dependence of the asymptotic form of the excited waves at late times is
through an exponential factor
\begin{displaymath}
\exp\bigl[\rmn{i} k_y (y + q\Omega t x) + \rmn{i} k_x x\bigr],
\end{displaymath}
where $k_y$ is the constant azimuthal wavenumber and $k_x$ is the radial
wavenumber at $t=0$. This latter quantity can be removed by redefining the
origin of time to correspond to the time at which the wave swings from leading
to trailing. Thus without loss of generality we may take $k_x$ to be zero.

When we go on to study the nonlinear development of the waves, it is natural
to consider disturbances with the same coordinate dependence as described
above. It is accordingly convenient to transform to shearing coordinates
$(x',y')$, which are related to $(x,y)$ through
\begin{equation}
  x' = x,\ y' = y + q\Omega x t.
\end{equation}
The shearing coordinate $y'$ specifies the coordinate of an unperturbed fluid
element in the direction of the background flow, being $y$ at the reference
time $t=0$. It is also convenient to work in terms of the velocity
perturbation to the background shear
\mbox{$\vec{u}=\vec{v}+q\Omega{}x\vec{e}_y$}. Then, in shearing coordinates,
the coordinate dependence is on $y'$ alone, so that equations
(\ref{eq:continuity}) and (\ref{eq:motion}) become
\begin{equation}
  \label{eq:continuity-sc}
  \pderc{\rho}{t}{y'}
  + \pder{}{y'}\Bigl[\rho(u_y + q\Omega t u_x)\Bigr] = 0 
\end{equation}
and
\begin{eqnarray}
  \label{eq:motion-sc}
  \pderc{\vec{u}}{t}{y'} + (u_y + q\Omega t u_x)\pder{\vec{u}}{y'}
  + 2\vec{\Omega}\times\vec{u} - q\Omega u_x\vec{e}_y = \nonumber\\
  - \left(\frac{\vec{e}_y + q\Omega t\vec{e}_x}{\rho}\right)\pder{P}{y'}
  + \vec{f},
\end{eqnarray}
where we have indicated that time derivatives are to be taken at constant
$y'$.

\subsection{Lagrangian description}\label{sec:lagrange}

Equations (\ref{eq:continuity-sc}) and (\ref{eq:motion-sc}) describe a form of
gas dynamics in one spatial dimension. For such problems a simplification can
often be made by adopting a Lagrangian description that uses a spatial
coordinate that remains fixed on a fluid element. Such a coordinate,
$y_0(y',t)$, may be defined through the equation
\begin{equation}
  \label{eq:y0}
  \pderc{y'}{t}{y_0} = u_y + q\Omega t u_x.
\end{equation}
Following \citet{LyndenBellOstriker1967}, we introduce an undisturbed 'ghost'
flow, which in the present case is simply the background shear flow. We
consider the mapping of fluid elements from the disturbed flow to the ghost
flow and take $y_0$ to be the initial value of $y'$ for the corresponding
element of the ghost flow. We remark that with this specification we \emph{do
not} require that $y'$ for the disturbed flow and $y_0$ coincide at time $t=0$
which in turn allows for a non zero density perturbation at $t=0$. Note that
(\ref{eq:y0}) states that $y'$ moves with the component of fluid velocity
normal to the phase surfaces of constant $y'$.

It is convenient to work in terms of the specific volume $V=1/\rho$.
Conversion of the spatial variable from $y'$ to $y_0$ is effected by using the
relation
\begin{equation}
  \pder{y'}{y_0} = \frac{\rho_0}{\rho}= \frac{V}{V_0},
\end{equation}
which follows from (\ref{eq:continuity-sc}) and (\ref{eq:y0}). Here, $\rho_0$
is the uniform background density, which is fixed for a given fluid element
and $V_0=1/\rho_0$. In the Lagrangian description equations
(\ref{eq:continuity-sc}) and (\ref{eq:motion-sc}) thus transform to
\begin{equation}
  \label{eq:continuity-sc0}
  \pderc{V}{t}{y_0} = V_0\pder{(u_y + q\Omega t u_x)}{y_0},
\end{equation} 
\begin{equation}
  \label{eq:motion-x-sc0}
  \pderc{u_x}{t}{y_0} - 2\Omega u_y = -q\Omega t V_0\pder{P}{y_0} + f_x,
\end{equation}
\begin{equation}
  \label{eq:motion-y-sc0}
  \pderc{u_y}{t}{y_0} + (2-q)\Omega u_x = -V_0\pder{P}{y_0} + f_y.
\end{equation}

\subsection{Vortensity conservation}\label{sec:vortensity}

An equation for the evolution of the ratio of the vertical component of
vorticity to the density, which we refer to as the `vortensity', is obtained
from equations (\ref{eq:motion-x-sc0}) and (\ref{eq:motion-y-sc0}) by
multiplying the latter by $q\Omega t$ and subtracting it from the former. With
the help of equation (\ref{eq:continuity-sc0}) we then obtain 
\begin{equation}
  \label{eq:vortensity}
  \pder{Q}{t} + V_0\pder{(f_x - q\Omega t f_y)}{y_0} = 0,
\end{equation}
where the vortensity is given by
\begin{equation}
  Q = (2-q)\Omega V + V_0\pder{(q\Omega t u_y - u_x)}{y_0},
\end{equation}
and from now on we omit to indicate that time derivatives are taken at
constant $y_0$. We also find it convenient to separate specific volume and the
vortensity into background and perturbed parts according to
\mbox{$V=V_0+\delta{}V$} and \mbox{$Q=Q_0+\delta{}Q$}, respectively, where
\mbox{$Q_0=(2-q)\Omega/\rho_0$} is the background vortensity and
\begin{equation}
  \label{eq:delta-Q}
  \delta Q = (2-q)\Omega\delta V + V_0\pder{(q\Omega t u_y - u_x)}{y_0}
\end{equation}
is the vortensity perturbation. We remark that in the inviscid case, equation
(\ref{eq:vortensity}) shows that vortensity is strictly conserved on fluid
elements. This remains true when viscous forces arise from a bulk viscosity,
as can be easily seen by noting that in this case the viscous forces can be
taken into account as an addition to the pressure. Furthermore, equation
(\ref{eq:vortensity}) is in conservation law form. In Eulerian form this reads
\begin{displaymath}
  \pderc{\left(\rho Q\right)}{t}{y'} +
  {\partial \over \partial y'}
  \bigl[\rho Q(u_y + q\Omega t u_x) + (f_x - q\Omega t f_y)\bigr] = 0.
\end{displaymath}
It implies that vortensity is conserved across shocks for any kind of small
viscosity (even though it may not be conserved in passing trough the shock
width). Thus for infinitesimal viscosity we may adopt \mbox{$\delta{}Q=0$}
everywhere and apply the standard Rankine-Hugoniot conditions to determine the
changes to other quantities on passing through shocks. Note that this is only
valid for planar shocks, which we are considering here. Curved shocks are in
general associated with vortensity production \citep[see e.g.][]{Hayes1957}.

In this context we note that in the linear theory of wave excitation described
in \citet{HeinemannPapaloizou2009a}, vortensity perturbations in the form of
stationary waves play a vital role in that excitation of traveling spiral
density waves only occurs if such perturbations are present. Wave excitation
happens during a narrow time interval around $t=0$ (the time of the swing), at
which point a pair of counter-propagating spiral density waves linearly
couples to a stationary vortical wave. However, this coupling is not effective
at later times due to an increasing frequency mismatch between the vortical
wave and the spiral density waves, so that we may take $\delta Q=0$ when
describing the dynamics of the latter at late times.

\subsection{Conservation of wave momentum in the nonlinear regime}
\label{sec:wave-momentum}

It is also possible to formulate the conservation of wave momentum for motions
governed by equations (\ref{eq:continuity-sc0}) to (\ref{eq:motion-y-sc0}). To
do this we introduce the Lagrangian displacement through
\begin{equation}
  \pder{\vec{\xi}}{t} = -q\Omega\xi_x\vec{e}_y + \vec{u}.
\end{equation}
The Lagrangian displacement $\vec{\xi}=(\xi_x,\xi_y)$ is measured relative to
the background or `ghost' fluid introduced in Section~\ref{sec:lagrange} and
is not necessarily small. In terms of it, equations (\ref{eq:motion-x-sc0})
and (\ref{eq:motion-y-sc0}) are given by
\begin{equation}
  \pdder{\xi_x}{t} - 2\Omega\pder{\xi_y}{t} - 2q\Omega^2\xi_x =
  -q\Omega t V_0\pder{P}{y_0} + f_x,
\end{equation}
\begin{equation}
  \pdder{\xi_y}{t} + 2\Omega\pder{\xi_x}{t} = -V_0\pder{P}{y_0} + f_y,
\end{equation}
and a time integration of equation (\ref{eq:continuity-sc0}) yields
\begin{equation}
  \delta V = V_0\pder{(\xi_y + q\Omega t\xi_x)}{y_0}.
\end{equation}

Having expressed the equations of motion in terms of Lagrangian displacement,
we can now derive a conservation law for wave momentum in the form of
\begin{equation}
  \label{eq:wave-momentum}
  \pder{\mathcal{U}}{t} + \pder{\mathcal{F}}{y_0} =
  -\vec{f}\cdot\pder{\vec{\xi}}{y_0},
\end{equation}
where the wave momentum density is
\begin{equation}
  \mathcal{U} = -\left(\pder{\vec{\xi}}{t}
  + \vec{\Omega}\times\vec{\xi}\right)\cdot\pder{\vec{\xi}}{y_0}
\end{equation}
and the associated flux is
\begin{equation}
  \mathcal{F} = -P\delta V + \mathcal{L}.
\end{equation}
Here, $\mathcal{L}$ is the Lagrangian density for the system in the absence of
dissipative forces. This is given by
\begin{equation}
  \mathcal{L} = \frac{1}{2}\pder{\vec{\xi}}{t}\cdot\pder{\vec{\xi}}{t}
  + (\vec{\Omega}\times\vec{\xi})\cdot\pder{\vec{\xi}}{t}
  + q\Omega^2\xi_x^2 - E,
\end{equation}
where $E(V,S)$ is the internal energy, to which the pressure is related by
\begin{equation}
  P = -\pderc{E}{V}{S}.
\end{equation}
When there are no dissipative forces, equation (\ref{eq:wave-momentum}) is a
strict conservation law that implies that the wave momentum flux $\mathcal{F}$
is constant in a steady state. We remark that this discussion does not
necessarily require a uniform background. It applies when $V_0$ is not
constant and to the adiabatic case with $S_0$ not constant. By considering the
effect of external forces the flux $\mathcal{F}$ is found to be a momentum
flux that converts to an angular momentum flux on multiplication by an assumed
radius of the center of the box.

When dissipation is included wave momentum is no longer conserved. From the
discussion in Section~\ref{sec:vortensity} it is apparent that we can
investigate the role of shocks by including a bulk viscosity. This can be done
by adding a component $\Pi$ to the pressure given by
\begin{equation}
  \label{eq:bulk-viscosity}
  \Pi = -\frac{\zeta}{V^2}\frac {\partial V}{\partial t},
\end{equation}
where $\zeta$ is the coefficient of bulk viscosity. In this form it is clear
that this adds a positive contribution to the pressure when the gas is being
compressed, as is required to support shocks. The viscous force corresponding
to (\ref{eq:bulk-viscosity}) is given by
\begin{equation}
  \label{eq:pseudo-viscous-force}
  \vec{f} = -(q\Omega t\vec{e}_x + \vec{e}_y)V_0\pder{\Pi}{y_0}.
\end{equation}
When a bulk viscosity is included in this way, equation
(\ref{eq:wave-momentum}) is modified to read
\begin{equation}
  \label{eq:wame-momentum-visc}
  \pder{\mathcal{U}}{t} + \pder{(\mathcal{F} - \Pi\delta V)}{y_0} =
  -\Pi\pder{V}{y_0}.
\end{equation}
Thus, wave momentum is no longer conserved.

The right hand side of (\ref{eq:wame-momentum-visc}) is in fact related to the
local rate of dissipation of energy per unit mass, given by
\begin{equation}
  \epsilon = -\Pi\pder{V}{t}.
\end{equation}
Accordingly, the right hand side of equation (\ref{eq:wame-momentum-visc}) may
be re-expressed as
$\epsilon(\partial{}V/\partial{}t)^{-1}\partial{}V/\partial{}y_0$. For a
non-dispersive traveling wave, the quantity multiplying $\epsilon$ is simply
$-1/v_0$, where $v_0$ is the wave velocity in the shearing coordinate $y_0$.
Let us consider an outward (inward) propagating trailing wave, traveling in
the direction of increasing (decreasing) $y_0$, for which $v_0>0$ ($v_0<0$).
As is well known, such a wave carries a positive (negative) amount of
momentum. Equation (\ref{eq:wame-momentum-visc}) thus implies that the
momentum content of each wave is lost to the background. This is consistent
with both waves being associated with an outward momentum flux.

\section{Weakly nonlinear theory}\label{sec:analytic}

Although we are considering sheared disturbances with a specific coordinate
dependence (see Section~\ref{sec:shearcoord}), we have so far not made any
further approximations. In this section we now restrict our consideration to
waves with finite but sufficiently small amplitude so that all terms of higher
than quadratic order in the wave amplitude can be neglected. We will also
generally consider the waves to be in the far trailing regime.

We start by noting that the number of dynamical equations that describe
nonlinear spiral density waves can be reduced to just a pair of equations by
invoking vortensity conservation. Indeed, we can dispose of the dynamical
equation for $u_y$ by using (\ref{eq:delta-Q}) to express this quantity in
terms of $u_x$, $\delta{}V$, and $\delta{}Q$. As we discussed in the end of
Section~\ref{sec:vortensity}, we may set the vortensity perturbation
$\delta{}Q=0$, in which case
\begin{equation}
  \label{eq:uy-vortcons}
  u_y = \frac{1}{q\Omega t}\left[u_x +
  \frac{(q-2)\Omega\,\partial_{y_0}^{-1}\delta V}{V_0}\right].
\end{equation}
Here, $\partial_{y0}^{-1}$ denotes the inverse of the partial derivative with
respect to $y_0$. An arbitrary function of time arising from the integration
with respect to $y_0$ is left unspecified at this point. We will later
determine this by enforcing momentum conservation (see
Section~\ref{sec:dispersion} below). A consequence of (\ref{eq:uy-vortcons})
is that in the limit of late times, the characteristic magnitude of $u_y$ is
smaller than that of $u_x$ by a factor of $q\Omega t$. In this limit, it is
convenient to adopt the quadratic time variable
\begin{equation}
  \tau = \frac{q\Omega t^2}{2},
\end{equation}
in terms of which the equations of motion governing the nonlinear waves are
given by
\begin{equation}
  \label{eq:continuity-tau}
  \pder{V}{\tau} =
  V_0\left(\pder{u_x}{y_0} +
  \frac{1}{q\Omega t}\pder{u_y}{y_0}\right)
\end{equation}
\begin{equation}
  \label{eq:motion-x-tau}
  \pder{u_x}{\tau} = \frac{V_0 c^2}{V^2}\pder{V}{y_0}
  + \frac{2\Omega u_y}{q\Omega t}
  - V_0\pder{\Pi}{y_0}.
\end{equation}
Here, we recall that recall that \mbox{$c^2=-V^2\partial{}P/\partial{}V$}
evaluated at constant entropy is the square of the sound speed and it is
understood that \mbox{$t=\sqrt{2\tau/q\Omega}$}.

In the limit of large $\tau$ and vanishing viscosity, equations
(\ref{eq:continuity-tau}) and (\ref{eq:motion-x-tau}) become equivalent to the
standard equations for gas dynamics. The terms involving $u_y$ lead to small
corrections that arise from the non-uniformity of the shearing medium through
which the waves propagate. In the linear theory given in
\citet{HeinemannPapaloizou2009a}, they are seen to induce slow amplitude and
phase changes occurring on a time scale long compared to the inverse wave
frequency. These effects appear at first order in a WKBJ treatment and we
shall treat additional changes arising from nonlinearity and viscosity on the
same footing below.

The asymptotic similarity to one-dimensional gas dynamics described in the
previous paragraph suggests that we describe the nonlinear waves using the
Riemann invariants
\begin{equation}
  R_\pm = u_x \pm \int^V_{V_0}\frac{c\,\rmn{d}V}{V},
\end{equation}
in terms of which the equations of motion are given by
\begin{equation}
  \pder{R_\pm}{\tau} = \pm\frac{cV_0}{V}
  \pder{}{y_0}\left(R_\pm + \frac{u_y}{q\Omega t}\right)
  + \frac{2\Omega u_y}{q\Omega t} - V_0\pder{\Pi}{y_0}.
\end{equation}
We note that this pair of equations is still exact. Independently of wave
amplitude, in the limit of both large $\tau$ and small $\zeta$, they have
solutions consisting of simple waves. These are such that the forward
propagating wave has \mbox{$R_{+}=0$} and the backward propagating wave has
\mbox{$R_{-}=0$}. The propagation speed $cV_0/V$ corresponds to a propagation
speed $c$ when measured with respect to the Eulerian coordinate $y'$. In the
linear approximation the general solution is a superposition of simple waves.

\subsection{An approximate wave equation governing weakly nonlinear waves}
\label{sec:wave-equation}

At leading order in WKBJ theory, the Riemann invariants satisfy first order
linear wave equations, according to which $R_\pm$ travel in opposite
directions at constant amplitude with speed $c_0$, being the sound speed $c$
evaluated for \mbox{$V=V_0$}. We consider modifications to the leading order
dynamics resulting from weak nonlinearity in the wave amplitudes $R_\pm$,
corrections resulting from the presence of $u_y$ at large $\tau$, as well as
corrections resulting from a small amount of viscosity. As the last two
corrections are intrinsically small, we only consider them to linear order.

To express quantities of interest in terms of the Riemann invariants, we note
that \mbox{$u_x=(R_{+}+R_{-})/2$} and
\mbox{$\int^V_{V_0}(c/V)\,\rmn{d}V=(R_{+}-R_{-})/2$}. From these two relations
it follows that if we set \mbox{$V=V_0+\delta{}V$}, then to first order in the
wave amplitude $\delta{}V$ we have
\mbox{$c_0\delta{}V/V_0\approx(R_{+}-R_{-})/2$}. Recalling that $c_0$ is the
sound speed evaluated for \mbox{$V=V_0$}, the propagation speed is, to the
same level of accuracy, given by
\begin{equation}
  \label{eq:wave-speed}
  \frac{cV_0}{V} \approx
  c_0 - \left(\frac{\delta V}{V_0}\right) c_1 \approx
  c_0 - \left(\frac{R_{+} - R_{-}}{2}\right)\frac{c_1}{c_0},
\end{equation}
where
\begin{equation}
  c_1 = -V_0^2\pder{(c/V)}{V}\Big|_{V=V_0}.
\end{equation}
Note that $c_1=c_0$ for an isothermal equation of state.

We focus on the forward propagating wave, which to leading order in WKBJ
theory is described by $R_{-}$ (a corresponding parallel analysis can be done
for the backward propagating wave). In the dynamical equation for $R_{-}$, the
terms containing $u_y$ and $\Pi$ (which, as remarked above, we consider to be
linear) as well as the weakly nonlinear contribution due to the variable wave
speed, given by (\ref{eq:wave-speed}), involve both $R_{-}$ \emph{and}
$R_{+}$. However, it is possible to argue that $R_{+}$ may be neglected in the
equation for $R_{-}$. We can see this from the following considerations.

Assuming that the forward propagating wave is dominated by $R_{-}$, we can
estimate $R_{+}$ from its dynamical equation, which has \emph{small} terms
proportional to $R_{-}$ as sources. In the rest frame of $R_{-}$, these source
terms appear to be of high frequency because $R_{+}$ travels in the opposite
directions. This leads to the conclusion that the induced $R_{+}$ will be
small compared to the original $R_{-}$ and so its contribution to the
\emph{small} terms in the equation for $R_{-}$ may be neglected. We note that
such a situation is common in weakly nonlinear hyperbolic systems, see
\citet{HunterKeller1983} for a general treatment.

With $R_{+}$ dropped and terms involving $u_y$ and $\Pi$ treated as linear,
the equation for $R_{-}$ becomes
\begin{eqnarray}
  \label{eq:R-minus}
  \pder{R_{-}}{\tau} +
  \left(c_0 + \frac{c_1}{2c_0}R_{-}\right)\pder{R_{-}}{y_0} =
  \hspace{3.0cm}\nonumber\\
  - \frac{1}{q\Omega t}
  \left(c_0\pder{u_y}{y_0} - 2\Omega u_y\right)
  - V_0\pder{\Pi}{y_0},
\end{eqnarray}
where $u_y$ and $\Pi$, respectively defined in (\ref{eq:uy-vortcons}) and
(\ref{eq:bulk-viscosity}), are to first order in small quantities given by
\begin{equation}
  \label{eq:uy-minus}
  u_y = \frac{1}{2q\Omega t}
  \Bigl[R_{-} + (2-q)\Omega(c_0\partial_{y_0})^{-1}R_{-}\Bigr]
\end{equation}
and
\begin{equation}
  \Pi = \frac{1}{2}(2q\Omega\tau)^{1/2}\frac{\zeta}{V_0 c_0}\pder{R_{-}}{\tau}.
\end{equation}
In the last expression for the viscous pressure, we consider the bulk
viscosity coefficient $\zeta$ to be a small quantity, in which case we may
replace \mbox{$\partial{}R_{-}/\partial\tau$} by
\mbox{$-c_0\partial{}R_{-}/\partial{}y_0$}.

The two terms on the right hand side of (\ref{eq:uy-minus}) give rise to
various, qualitatively different effects when this expression is substituted
into (\ref{eq:R-minus}). Both terms lead to slow changes in the wave
amplitude. The first term also leads to a modification of the propagation
speed. The second term also introduces dispersion. All of these effects are
consistent with linear WKBJ theory.

The change in the propagation speed is taken care of by transforming to a
frame moving with speed
\begin{displaymath}
  \left(1 + \frac{1}{4q\Omega\tau}\right) c_0
\end{displaymath}
in the $y_0$ direction. The new spatial coordinate is thus
\begin{equation}
  \eta = y_0 - c_0\int\left(1 + \frac{1}{4q\Omega\tau}\right)\rmn{d}\tau.
\end{equation}
Finally, for notational convenience, we introduce
\begin{equation}
U = \frac{c_1}{2c_0}R_{-}.
\end{equation}
The governing equation for waves propagating in the $y_0$ direction is then
\begin{eqnarray}
  \label{eq:U}
  \pder{U}{\tau} + U\pder{U}{\eta} - \frac{U}{4\tau} =
  \hspace{4.5cm}\nonumber\\
  \frac{\kappa^2(c_0\partial_\eta)^{-1}U}{4q\Omega\tau} +
  \frac{1}{2}(2q\Omega\tau)^{1/2}\pder{}{\eta}
  \left(\zeta\pder{U}{\eta}\right),
\end{eqnarray}
where it is understood that $\partial/\partial\tau$ is now to be evaluated at
constant $\eta$. At this point we remark that as indicated above, a parallel
analysis can be performed to obtain the corresponding equation for the
backward propagating wave described by \mbox{$U=c_1R_{+}/(2c_0)$}. This is the
same as (\ref{eq:U}), but with the sign of the dispersive term (first term on
the right hand side) reversed. Its solutions may be obtained from those of
(\ref{eq:U}) by use of the transformation \mbox{$\eta\rightarrow-\eta$},
\mbox{$y_0\rightarrow-y_0$}, \mbox{$U\rightarrow-U$}.

Note that the viscous term increases in magnitude as $\tau$ increases as a
consequence of the shear causing the radial wavelength of the disturbance to
shorten with time. The bulk viscosity coefficient $\zeta$ thus needs to
decrease with time if the relative importance of the viscous term is not to
change. The particular form of $\zeta$ that we use in
Section~\ref{sec:simulations} to integrate the equation of motion numerically
has in fact this property.

Equation (\ref{eq:U}) may be viewed as a modified form of Burgers equation.
The first two terms on the right hand side correctly account for first order
changes in amplitude and phase arising in WKBJ theory. The only nonlinearity
comes from the advection term on the left hand side, which causes wave
steepening and the formation of shocks.

\subsection{Nonlinear sawtooth waves}\label{sec:sawtooth}

It is possible to consider solutions of equation (\ref{eq:U}) with different
ordering regimes for the different terms. We have just discussed the case
where nonlinear, dispersive and dissipative effects are all small. Physically
this corresponds to a short wavelength wave in the linear regime with a
conserved wave momentum which has not had time enough to steepen and form
shocks. We now go on to consider the case when the nonlinear term dominates
the dispersive and dissipative terms corresponding to a situation where the
wave is in the process of forming a shock. Setting
\mbox{$\bar{U}=U/\tau^{1/4}$}, adopting a new time coordinate
\mbox{$\bar{\tau}=(4/5)\,\tau^{5/4}$} and neglecting dispersive and
dissipative terms on the right hand side of (\ref{eq:U}), this equation is
transformed to the inviscid Burgers equation
\begin{equation}
  \label{eq:burgers}
  \pder{\bar{U}}{\bar{\tau}} + \bar{U}\pder{\bar{U}}{\eta} = 0.
\end{equation}

It is a well known result in nonlinear acoustics that an initially
monochromatic wave that satisfies Burgers equation steepens to form a sawtooth
wave \citep[see][and references therein]{Parker1992}. Although we do not begin
with a strictly monochromatic wave we are close to that situation and the
numerical simulations of Section~\ref{sec:simulations} show that the solutions
to the full nonlinear equations of motion do in fact approach a sawtooth wave,
so we shall consider these solutions here.

A sawtooth wave consists of a periodic array of shocks. Because we are
considering the inviscid form of Burgers equation, we are implicitly taking
the limit of vanishing viscosity. In this limit, the shocks in the sawtooth
wave are infinitely thin and may be treated as discontinuities. The change in
$\bar{U}$ is then determined from the conservation law associated with
(\ref{eq:burgers}). For a stationary (i.e.\ non-traveling) wave, this
requires that $\bar{U}$ changes sign across the shock.

The solution of (\ref{eq:burgers}) corresponding to a stationary sawtooth wave
with wavelength $\lambda$ can be expressed in terms of $U$ as 
\begin{equation}
  \label{eq:sawtooth}
  U = \frac{5\eta}{4\tau}\,\frac{1}{1 - (\tau_0/\tau)^{5/4}}
\end{equation}
where the wave is periodically extended for \mbox{$|\eta|>\lambda/2$} and
$\tau_0$ is an arbitrary integration constant. The wavelength $\lambda$ (which
we remark is time-independent when measured with respect to the shearing
coordinate $y_0$) may be specified arbitrarily (but see below). The solution
is linear in $\eta$ between a periodic array of shocks with amplitude
$\lambda{}U/\eta$.

Noting that \mbox{$u_x=c_0U/c_1$} and using the definitions of $\tau$ to
express the result in terms of the time $t$, we see that for the fundamental
period of the sawtooth wave, when $t$ and $\tau$ are large enough that
$\tau_0$ may be neglected, we have
\begin{equation}
  \label{eq:sawtooth-x}
  u_x = \frac{c_0}{2c_1}\frac{5\eta}{q\Omega t^2}.
\end{equation}
The maximum velocity amplitude is obtained by setting \mbox{$\eta=\lambda/2$}.
The wavelength $\lambda$ thus determines the shock amplitude at a given late
time. The fact that this amplitude decays implies dissipation of energy, which
may be interpreted as being due to the action of a vanishingly small bulk
viscosity within an infinitely thin shock layer. As we discussed in
Section~\ref{sec:wave-momentum}, dissipation of energy is always associated
with angular momentum transfer to the background.

\subsection{Correction for dispersion}\label{sec:dispersion}

We now estimate corrections to the sawtooth solution resulting from dispersion
but still take viscosity to be vanishingly small. We work in the limit of
large $\tau$ so that again $\tau_0$ in (\ref{eq:sawtooth}) is neglected. The
governing equation is equation (\ref{eq:U}) with \mbox{$\zeta=0$} which reads
\begin{equation}
  \label{eq:U-novisc}
  \pder{U}{\tau} + U\pder{U}{\eta} - \frac{U}{4\tau} =
  \frac{\kappa^2(c_0\partial_\eta)^{-1}U}{4q\Omega\tau}
\end{equation}

We begin by allowing for a small change to the time dependent propagation
speed of the wave, which we are regarding to be of the same order as the
correction we are estimating. This gives us the freedom to ensure that the
correct jump conditions across the shock are maintained. We thus add a term
\mbox{$c_2(\tau){\partial{}U}/{\partial\eta}$} to both sides of
(\ref{eq:U-novisc}), where $c_2(\tau)$ is the as yet undefined small increase
in the propagation speed. In order to deal with the additional term on the
left hand side, we redefine the co-moving coordinate $\eta$ so that it is
boosted by the speed $c_2(\tau)$ and thus given by
\begin{equation}
  \label{eq:comoving}
  \eta = y_0 - c_0\int\left(1 + \frac{1}{4q\Omega\tau}\right)\rmn{d}\tau
  - \int\! c_2(\tau)\,\rmn{d}\tau.
\end{equation}
In terms of the new co-moving coordinate, equation (\ref{eq:U-novisc}) becomes
\begin{equation}
  \label{eq:U-comoving}
  \pder{U}{\tau} + U\pder{U}{\eta} - \frac{U}{4\tau}=
  \frac{\kappa^2(c_0\partial_\eta)^{-1}U}{4q\Omega\tau}
  + c_2(\tau)\pder{U}{\eta},
\end{equation}
where we emphasize again that the last term on the right hand side is supposed
to lead to a small correction of the same order as that produced by the first
term on the right hand side.

To find an approximate solution we set \mbox{$U=U_0+U_1$} where $U_0$ is the
sawtooth solution given by (\ref{eq:sawtooth}) and $U_1$ is a small
correction. Dropping terms that are quadratic in $U_1$ on the left hand side
of (\ref{eq:U-comoving}), and neglecting $U_1$ on the right hand side,
we find that $U_1$ satisfies the equation
\begin{equation}
  \label{eq:U1}
  \pder{U_1}{\tau} + \frac{5\eta}{4\tau}\pder{U_1}{\eta} +
  \frac{U_1}{\tau} =
  \frac{\kappa^2}{4q\Omega\tau}\left[\frac{5\eta^2 + g(\tau)}{8c_0\tau}\right]
\end{equation}
Here, $g(\tau)$ is an arbitrary function arising partly from the operator
$\partial_\eta^{-1}$ and partly from the incorporation of the last term on the
right hand side of (\ref{eq:U-comoving}). It can be chosen to ensure that at
any time the wave has no mean momentum, i.e.
\begin{equation}
  \label{eq:constraint}
  \int_{-\lambda/2}^{\lambda/2}(U_0 + U_1)\,\rmn{d}\eta = 0.
\end{equation}
To solve (\ref{eq:U1}) we set
\mbox{$U_1(\eta,\tau)=\alpha(\tau)\eta^2+\beta(\tau)$}. The functions
$\alpha(\tau)$ and $\beta(\tau)$ can be found by substituting this ansatz into
(\ref{eq:U1}) and integrating with respect to $\tau$. In fact, one has only to
find $\alpha(\tau)$, since $\beta(\tau)$ can then be determined using the
constraint (\ref{eq:constraint}). The dispersive correction is thereby found
to be
\begin{equation}
  \label{eq:correction}
  U_1(\eta,\tau) = \frac{\kappa^2(\eta^2 - \lambda^2/12)}{16q\Omega\tau c_0}.
\end{equation}

We now turn to the evaluation of $c_2(\tau)$, which we introduced so as to
ensure that the correct shock jump conditions are satisfied. From
(\ref{eq:U-comoving}) it is apparent that
\begin{displaymath}
  U^2/2 - c_2 U \approx U_0^2/2 + U_0(U_1 - c_2)
\end{displaymath}
should be conserved when passing through the shock. Since $U_0$ changes sign
across the shock and $U_1$ does not, this leads to
\begin{equation}
  c_2(\tau) = U_1(\lambda/2,\tau) =
  \frac{\kappa^2 \lambda^2}{96q\Omega\tau c_0}.
\end{equation}
The corresponding correction to the co-moving coordinate defined in
(\ref{eq:comoving}) is thus proportional to $\ln\tau$ and hence small.

In order to measure the modification of the sawtooth form due to the
dispersive correction, we evaluate
\begin{equation}
  \mathcal{R} =
  \left(\frac{\int_{-\lambda/2}^{\lambda/2} U_1^2\,\rmn{d}\eta}
             {\int_{-\lambda/2}^{\lambda/2} U_0^2\,\rmn{d}\eta}\right)^{1/2} =
  \frac{\kappa^2\lambda}{20\sqrt{15}\,q\Omega c_0}.
\end{equation}
Interestingly, this expression does not depend on time but only on the
wavelength $\lambda$. The dependence on $\lambda$ is such that the dispersive
correction to the sawtooth is small for sufficiently short wavelengths.
However, even for $\lambda=2\pi{}c_0/\Omega$, which is the optimal wavelength
for excitation by vortensity fluctuations in a Keplerian disk
\citep{HeinemannPapaloizou2009a}, we find
$\mathcal{R}=\pi/15^{3/2}\approx{}0.054$, suggesting that dispersive
corrections to the sawtooth profile may be neglected in practice.

\section{Numerical Simulations}\label{sec:simulations}

We have obtained solutions of the set of nonlinear equations
(\ref{eq:continuity-sc0}) to (\ref{eq:motion-y-sc0}) by means of numerical
simulations. In the limit \mbox{$t\rightarrow\infty$} this set of equations
leads to a pair of equations that resembles the standard equations for one
dimensional inviscid gas dynamics with \mbox{$\tau=q\Omega{}t^2/2$} as the
effective time variable. As with those, we expect nonlinear wave steepening
leading to the formation shock waves \citep[e.g.][]{LandauLifshitz1987}. In
order to represent these, viscous dissipation must be included. Although this
has to be significant only in the thin transition region between the pre-shock
and post-shock fluid. Here we deal with this by adopting the procedure of
\citet{VonNeumannRichtmyer1950} in which an artificial viscous pressure,
$\Pi$, is added to the gas pressure, $P$. For the simulations presented here
we adopt an isothermal equation of state, such that \mbox{$P=c^2/V$}, with the
isothermal sound speed $c=c_0$ being constant. Then equations
(\ref{eq:continuity-sc0}) to (\ref{eq:motion-y-sc0}) together with
(\ref{eq:pseudo-viscous-force} become
\begin{equation}
  \label{eq:continuity-num}
  \rho_0\pder{V}{t} = \pder{(u_y + q\Omega t u_x)}{y_0},
\end{equation} 
\begin{equation}
  \rho_0\left[\pder{u_x}{t} - 2\Omega u_y\right] =
  -q\Omega t \pder{(c^2/V + \Pi)}{y_0},
\end{equation}
\begin{equation}
  \label{eq:motion-y-num}
  \rho_0\left[\pder{u_y}{t} + (2-q)\Omega u_x\right] =
  -\pder{(c^2/V + \Pi)}{y_0}.
\end{equation}
The form of the viscous pressure $\Pi$ is given by (\ref{eq:bulk-viscosity})
as
\begin{equation}
  \label{eq:Pi}
  \Pi = -\frac{\zeta}{V^2}\pder{V}{t},
\end{equation}
where, following \citet{VonNeumannRichtmyer1950}, we take the bulk viscosity
coefficient to be
\begin{equation}
  \label{eq:zeta}
  \zeta = (\rho_0\beta\Delta y_0)^2 V\left|\pder{V}{t}\right|.
\end{equation}
Here, $\Delta{}y_0$ is the computational grid spacing in $y_0$ and $\beta$ is
a constant of order unity which can be adjusted to smear shocks over a fixed
number of numerical grid points (typically between 3 and 5). We adopted
\mbox{$\beta=3$} in our simulations. 

We remark that because $y_0$ is a shearing coordinate, shock widths measured
with respect to $y_0$ increase linearly with time at late times. This can be
seen as follows. Inserting (\ref{eq:zeta}) into (\ref{eq:Pi}), and using the
fact that for an outgoing wave we have to lowest order
\begin{equation}
  \pder{V}{t} = -q\Omega tc\pder{V}{y_0}
\end{equation}
(see Section~\ref{sec:wave-equation}), we find
\begin{equation}
  \Pi = -\frac{(\rho_0 c\beta q\Omega t\Delta y_0)^2}{V}
  \pder{V}{y_0}\left|\pder{V}{y_0}\right|.
\end{equation}
This expression has an additional factor $(q\Omega t)^2$ as compared to a
standard case in non shearing coordinates. Thus the shock width scales as
\mbox{$\beta{}q\Omega{}t\Delta{}y_0$}, which indeed increases linearly with
time. Note, however, that \mbox{$y_0\sim{}q\Omega{}tx$} at late times, which
means that the width of a shock remains fixed when measured with respect to
$x$. Thus, an observer in the ``unsheared'' coordinate frame $(x,y)$ will, at
late times, see a shock of constant width propagating in the $x$-direction.

In practice, artificial viscosity is only significant in regions of strong
compression, where \mbox{$\partial{}V/\partial{}t<0$}. Thus, the viscous
pressure $\Pi$ is often set to zero in regions of rarefaction, where
\mbox{$\partial{}V/\partial{}t>0$}. In our simulations, steep gradients never
occur in these latter regions, so that this makes little difference, and we
have found empirically that slightly better results are obtained without this
modification. 

Simulations were carried out on an equally spaced computational grid in $y_0$
with typically 16384 grid points. We note that because of this rather high
resolution, the increase of the shock widths as discussed above is not
actually visible in the plots shown below. The computational domain was taken
to be periodic with period $2\pi{}c/\Omega$, which equals the optimal wave
length for wave excitation for a Keplerian rotation profile
\citep{HeinemannPapaloizou2009a}. In order to discretize the equations of
motion, we have adopted the same staggered leapfrog scheme that was used in
\citet{VonNeumannRichtmyer1950}. This scheme is second order accurate in space
and time away from shocks, but effectively becomes only first order accurate
(in space) in their vicinity.

\subsection{Initial conditions}

We start the numerical integration of (\ref{eq:continuity-num}) to
(\ref{eq:motion-y-num}) at a time corresponding to a couple of orbits before
the swing. The initial condition consists of a stationary vortical wave of
small amplitude. To derive this initial condition, we note that in the linear
inviscid regime, this system of equations is -- according to
\citet{HeinemannPapaloizou2009a} -- equivalent to the inhomogeneous
second-order wave equations
\begin{eqnarray}
  \label{eq:linear-wave-equation-ux-dV}
  \left[\pdder{}{t} + (q^2 \Omega^2 t^2 + 1)k^2 c^2 + \kappa^2
  \pm 2q\Omega\rmn{i}kc\right]
  \left(\frac{u_x}{c}\mp\frac{\delta V}{V_0}\right)
  \nonumber\\
  = \left(\rmn{i}k c\mp 2\Omega\right)\rho_0\delta Q
\end{eqnarray}
and
\begin{equation}
  \label{eq:linear-wave-equation-uy}
  \left[\pdder{}{t} + (q^2 \Omega^2 t^2 + 1)k^2 c^2 +
  \kappa^2\right]\frac{u_y}{c}
  = -q\Omega t\rmn{i} k_y c\,\rho_0\delta Q,
\end{equation}
where $\delta{}V$, $u_x$, $u_y$, and the vortensity perturbation $\delta{}Q$
are assumed to vary harmonically in space as $\exp(\rmn{i}k{}y_0)$. We note
that vortensity conservation implies that $\delta{}Q$ does not vary in time at
linear order.

At early times before the swing, we may drop the double time derivative in
(\ref{eq:linear-wave-equation-ux-dV}) and (\ref{eq:linear-wave-equation-uy})
to obtain the slowly varying vortical wave solutions
\begin{equation}
  \label{eq:vortical-wave-ux-dV}
  \left(\frac{u_x}{c}\mp\frac{\delta V}{V_0}\right) =
  \frac{\left(\rmn{i}k c\mp 2\Omega\right)\rho_0\delta Q}
  {(q^2 \Omega^2 t^2 + 1)k^2 c^2 + \kappa^2 \pm 2q\Omega\rmn{i}kc}
\end{equation}
and
\begin{equation}
  \label{eq:vortical-wave-uy}
  \frac{u_y}{c} = -\frac{q\Omega t\rmn{i}k_y c\,\rho_0\delta Q}
  {(q^2 \Omega^2 t^2 + 1)k^2 c^2 + \kappa^2},
\end{equation}
where it is understood that the real part of the right hand sides is to be
taken.

We initialize the integration with the vortical wave solution
(\ref{eq:vortical-wave-ux-dV}) and (\ref{eq:vortical-wave-uy}) at
\mbox{$t=-4\pi/\Omega$}, corresponding to two orbits before the swing. In this
way we generate initial conditions that lead to a time dependent evolution
that firstly reproduces the linear wave excitation phase described in
\citet{HeinemannPapaloizou2009a}, in which the vortical wave gives rise to the
excitation of a pair of counter-propagating spiral density waves at
\mbox{$t=0$}. Subsequent evolution causes the spiral density waves to enter
the nonlinear regime considered in this paper and described analytically in
Section~\ref{sec:analytic} above.

We assume a Keplerian rotation profile by setting \mbox{$q=3/2$}. For the
azimuthal wavenumber we choose \mbox{$k=1/H$}, where \mbox{$H=c/\Omega$} is
the putative density scale height. This corresponds to an azimuthal wavelength
of $2\pi{}H$, which is the optimal wavelength for spiral density wave
excitation \citep[see][for details]{HeinemannPapaloizou2009a}. The amplitude
of the vortensity perturbation $\delta{}Q$ we take to be $0.1\Omega/\rho_0$.

\subsection{Numerical results}\label{sec:numerical-results}

\begin{figure}
  \centerline{\includegraphics{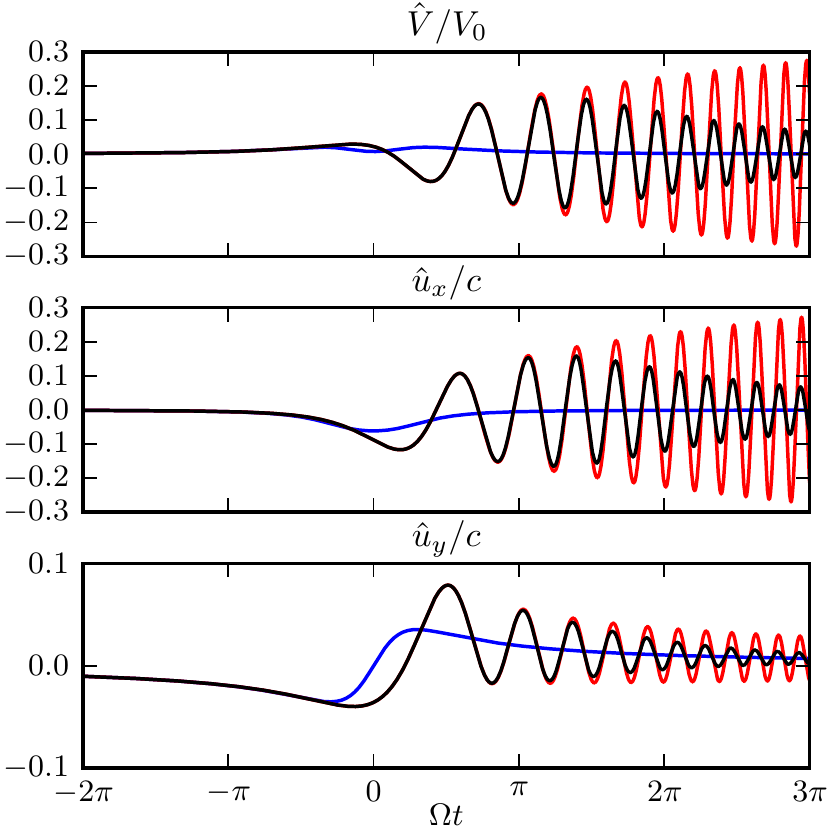}}
  \caption{Evolution of the \mbox{$k=1/H$} mode amplitudes according to the
  linear (red, attaining the largest values at late times) and nonlinear
  (black) equations of motion. The blue curves represent the corresponding
  vortical wave solutions (\ref{eq:vortical-wave-ux-dV}) and
  (\ref{eq:vortical-wave-uy}). The linear solution was obtained by integrating
  the linearized equations of motion using a high order Runge-Kutta method.}
  \label{fig:nonlinear-amplitudes}
\end{figure}

In Figure~\ref{fig:nonlinear-amplitudes} we compare the time evolution of
solutions to the nonlinear equations of motion (\ref{eq:continuity-num}) to
(\ref{eq:motion-y-num}) obtained using the artificial viscosity method with
solutions of the linear wave equations (\ref{eq:linear-wave-equation-ux-dV})
and (\ref{eq:linear-wave-equation-uy}). To make this comparison, we computed
the $k=1/H$ Fourier components of the nonlinear solutions for $u_x$, $u_y$,
and $\delta{}V$ (denoted respectively by $\hat{u}_x$, $\hat{u}_y$, and
$\delta\hat{V}$), and plotted the result against the linear solutions.

The linear and nonlinear solutions are seen to agree closely during the
leading phase \mbox{$t<0$} and also during the early trailing phase
\mbox{$t>0$} after the wave excitation has taken place. This is consistent
with the fact that the amplitude of the vortensity perturbation $\delta Q$ was
chosen small enough to make the excitation process linear. Nonlinearity occurs
later as the waves propagate. This is on account of the increasing amplitude
implied as a consequence of wave momentum conservation (see
Section~\ref{sec:wave-momentum} above). A consequence of nonlinearity and
shock development is that the Fourier amplitudes $\hat{u}_x$ and
$\delta\hat{V}$ do not continue to increase with time as predicted by the
linear theory but instead attain maximum values and then decay. The maximum
relative density perturbation is roughly 20\%, which is attained at
\mbox{$t\sim{}3\pi/4\Omega$}, is similar to that seen in nonlinear simulations
of the MRI \citep{HeinemannPapaloizou2009b}. It is interesting to note that
although the nonlinear waves decay while the linear waves increase in
amplitude they maintain the same phases at a given time. We will comment on
this further below in Section~\ref{sec:propagation}.

\begin{figure}
  \centerline{\includegraphics{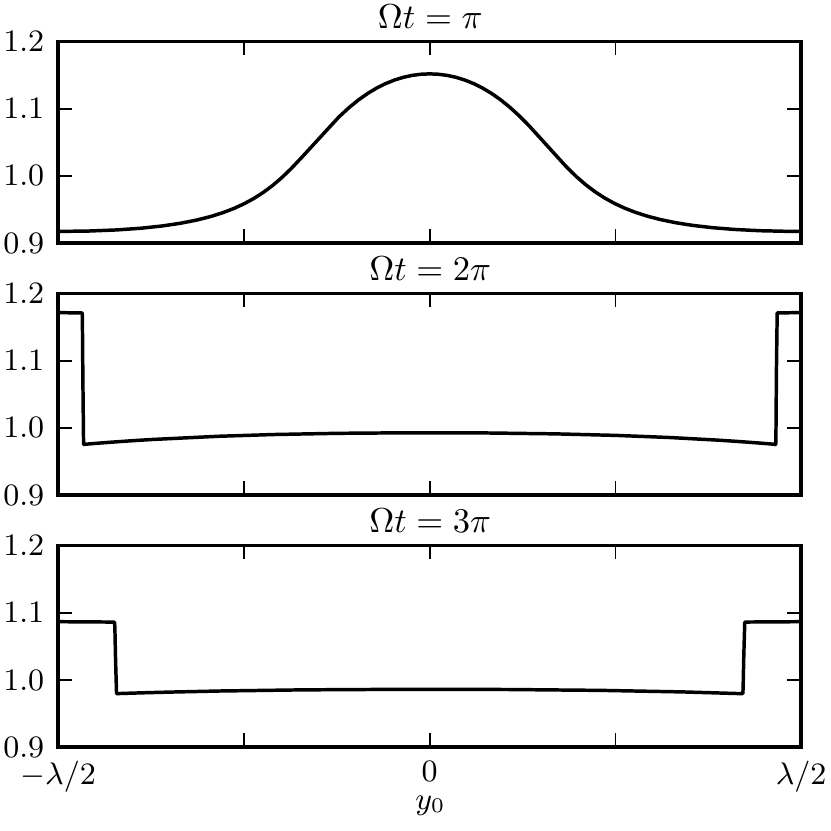}}
  \caption{Spatial profile of the specific volume at different times in the
  trailing phase.}
  \label{fig:nonlinear-evolution}
\end{figure}

The decay of the nonlinear wave amplitudes is due to a transfer of wave energy
to smaller scales due to nonlinear mode coupling. That such energy transfer
indeed occurs is illustrated in Figure~\ref{fig:nonlinear-evolution} where we
plot the specific volume $V$ as a function of $y_0$, over one wavelength, at
three different times in the trailing phase. Already half an orbit after
excitation, nonlinear steepening has lead to a notable distortion of the
initial harmonic profile. After one orbit, further steepening has resulted in
the formation of two shock fronts corresponding to the forward and backward
propagating waves. These separate regions of almost constant specific volume.
At this time the two shocks travel in opposite directions away from the
central region. Yet another half of an orbit later, the spatial profile
remains essentially unchanged, but the shock amplitudes have decreased by
approximately a factor of two because of the energy dissipation associated
with them.

\begin{figure}
  \centerline{\includegraphics{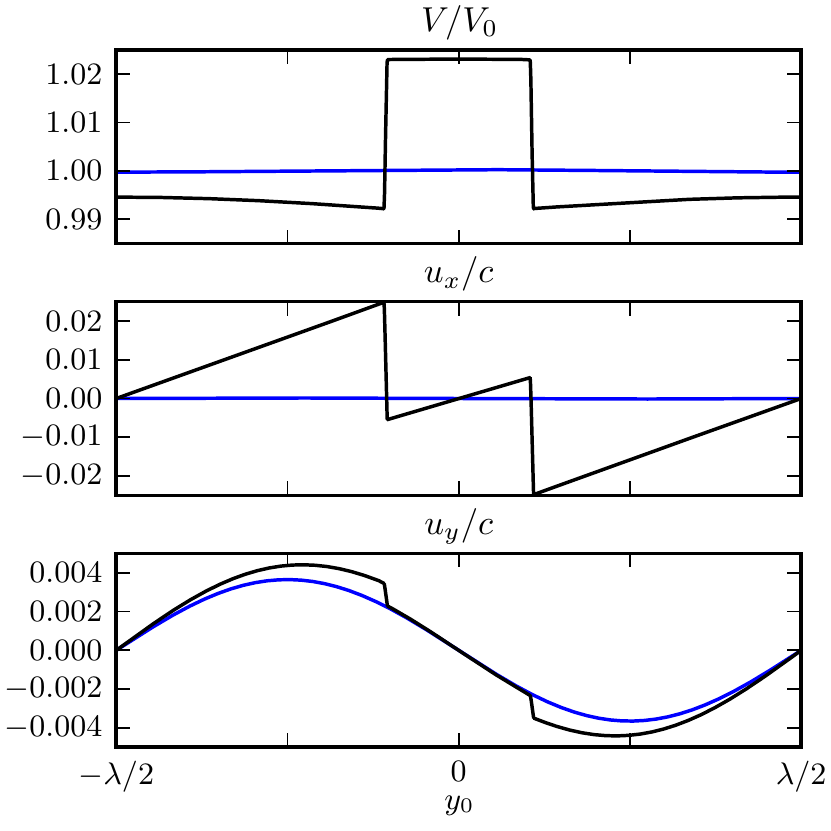}}
  \caption{Spatial profiles of all fluid variables roughly three orbits after
  excitation (\mbox{$\Omega{}t\approx6\pi$}). The black curve corresponds to
  the numerical solution. Also shown in blue are the corresponding vortical
  wave solutions (\ref{eq:vortical-wave-ux-dV}) and
  (\ref{eq:vortical-wave-uy}).}
  \label{fig:nonlinear-evolution-late}
\end{figure}

The spatial profiles of the velocity components and the specific volume
perturbation three orbits after excitation are plotted in
Figure~\ref{fig:nonlinear-evolution-late}. The shocks in the radial velocity
$u_x$ are seen to be as pronounced as those in the relative specific volume
perturbation $\delta{}V/V_0$. In this case they separate regions where the
velocity varies almost linearly with $y_0$, corresponding to a sawtooth form.
In comparison to $\delta{}V$ and $u_x$, the spatial variation in the azimuthal
velocity field $u_y$ remains predominantly harmonic. As indicated in
Figure~\ref{fig:nonlinear-evolution-late}, this sinusoidal variation
corresponds to the non-oscillatory vortical wave solution that follows from
(\ref{eq:vortical-wave-uy}). In contrast to this, the corresponding amplitudes
obtained from (\ref{eq:vortical-wave-ux-dV}) lead to negligibly small values
of $\delta{}V/V_0$ and $u_x/c$. This suggests that in the case of $u_y$, the
shock amplitude decays faster than the vortical wave amplitude, whereas the
opposite seems to be the case for $\delta{}V$ and $u_x$. This is discussed
further below.

\subsection{Nonlinear damping}\label{sec:damping}

We note that when a forward propagating wave has attained a sawtooth form,
equation (\ref{eq:sawtooth-x}), with \mbox{$c_0=c_1=c$} which applies to the
numerical simulations, gives $u_x$ as a function of the co-moving coordinate
$\eta$ with a periodic extension for \mbox{$|\eta|>\lambda/2$}, with $\lambda$
being the azimuthal wave length of the original linear wave. The shock
discontinuities are accordingly located at $\eta=(n+1/2)\lambda$, where
\mbox{$n{}\in\field{Z}$} and the jump across the shock is given by
\begin{equation}
  \label{eq:jump}
  \Delta (u_x) = \frac{5\lambda}{2q\Omega t^2}.
\end{equation}
This applies to waves traveling in either direction as neither is preferred.
Thus weakly nonlinear shock jumps decay quadratically in time. The decay of
the shock amplitudes necessarily leads to a decay of the fluctuation energy
per unit mass, defined as
\begin{equation}
  \label{eq:fluctuation-energy}
  \mathcal{E} = \frac {u_x^2 + u_y^2}{2}
  + \frac{c^2}{V_0^2}\frac{\delta V^2}{2}.
\end{equation}
Because for weakly nonlinear waves propagating in one direction, one of the
Riemann invariants propagates while the other is constant, it follows that
(\ref{eq:jump}) gives the magnitude of the shock jump for both $u_x$ and
$c\delta{}V/V_0$. From vortensity conservation, it follows that $u_y$ decays
more rapidly than $\delta{}V$ and $u_x$ by one power of $t$ (see
discussion in Section~\ref{sec:analytic} above) and so may be dropped from
(\ref{eq:fluctuation-energy}) as \mbox{$t\rightarrow\infty$}. In this limit
the fluctuation energy averaged over a wavelength associated with a pair of
shock waves is given by
\begin{equation}
  \label{eq:asymptotic-energy}
  \langle\mathcal{E}\rangle =
  \frac{1}{6}\left(\frac{5\lambda}{2q\Omega t^2}\right)^2.
\end{equation}
Thus the fluctuation energy at late time decays as $t^{-4}$.

\begin{figure}
  \centerline{\includegraphics{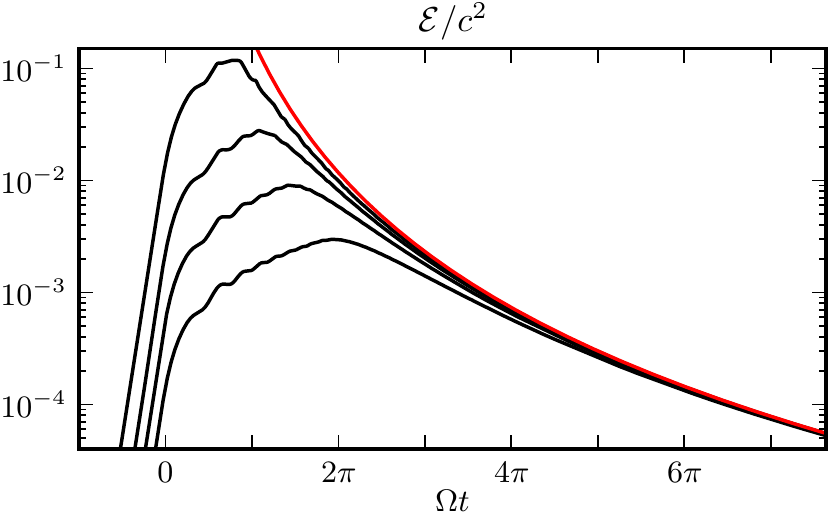}}
  \caption{Time evolution of the fluctuation energy per unit mass
  $\mathcal{E}$, as defined by equation (\ref{eq:fluctuation-energy}). Black
  curves not uppermost: Energy obtained numerically using the artificial
  viscosity method. Red and uppermost curve: Energy obtained from equation
  (\ref{eq:asymptotic-energy}) obtained from weakly nonlinear theory, see
  Section~\ref{sec:damping}.}
  \label{fig:nonlinear-energy}
\end{figure}

We note that in the limit \mbox{$t\rightarrow\infty$}, the above description
is ultimately independent of the initial wave amplitude. We have tested this
inherently nonlinear expectation against the behavior of our numerical
simulations. In Figure~\ref{fig:nonlinear-energy} we plot the fluctuation
energy (\ref{eq:fluctuation-energy}) for waves initiated with four different
amplitudes of the vortensity perturbation $\delta{}Q$. According to the linear
excitation mechanism described by \citet{HeinemannPapaloizou2009a}, this
amplitude, determines the amplitude of the excited linear wave. For the
examples illustrated the fluctuation energies obtained shortly after
excitation differ by almost two orders of magnitude. However, as $t$
approaches large values, the fluctuation energies associated with the
numerical solutions all approach the predicted late-time fluctuation energy
(\ref{eq:asymptotic-energy}). This is because the more the initially linear
wave is amplified, the sooner it becomes nonlinear, and larger amplitude waves
decay faster than smaller amplitude ones.

\subsection{Shock propagation}\label{sec:propagation}

In this section we will investigate the space-time dependence of the nonlinear
waves. We remind the reader that for the forward propagating weakly nonlinear
solution described in Section~\ref{sec:wave-equation}, the co-moving
coordinate is given by (\ref{eq:comoving}) as
\begin{equation}
  \label{eq:comoving-reminder}
  \eta = y_0 - c\tau - \frac{c\ln|\tau|}{4q\Omega}
  \left(1+\frac{\kappa^2\lambda^2}{24c^2}\right),
\end{equation}
where we have now reverted back to quadratic time variable $\tau$ used in
Section~\ref{sec:wave-equation}. For the backward propagating wave $c$ is
replaced by $-c$ in (\ref{eq:comoving-reminder}). Apart from asymptotically
insignificant amplitude (through $\lambda$) and logarithmic corrections, the
functional form of the shock wave solution is the product of a time-dependent
amplitude and a function of the phase \mbox{$y_0-c\tau$} as is also the case
for linear waves. Thus the speed at which the shock fronts propagate is
approximately equal to the sound speed, being the phase speed of linear waves
in the far trailing regime.

We are now in a position to understand the evolution of individual Fourier
harmonics as obtained from the numerical solution of the nonlinear equations,
as shown in Figure~\ref{fig:nonlinear-amplitudes}. There we see that while
nonlinear effects do lead to a decrease in the oscillation amplitude, they do
not seem to alter the oscillation period. This follows immediately from the
fact that the linear and nonlinear waves maintain approximately the same phase
as a function of time.

\subsection{Shock wave profile}

In Section~\ref{sec:numerical-results} we illustrated the spatial profile of a
pair of shock waves observed in a numerical simulation. This is characterized
by a top-hat profile in the specific volume perturbation and a double sawtooth
profile in the radial velocity. A similar double sawtooth profile is also seen
in the azimuthal velocity, but there it is disguised by the spatially
sinusoidal vorticity perturbation that we used as initial condition. As we
will now demonstrate, all of these features can be understood from weakly
nonlinear theory.

\begin{figure}
  \centerline{\includegraphics{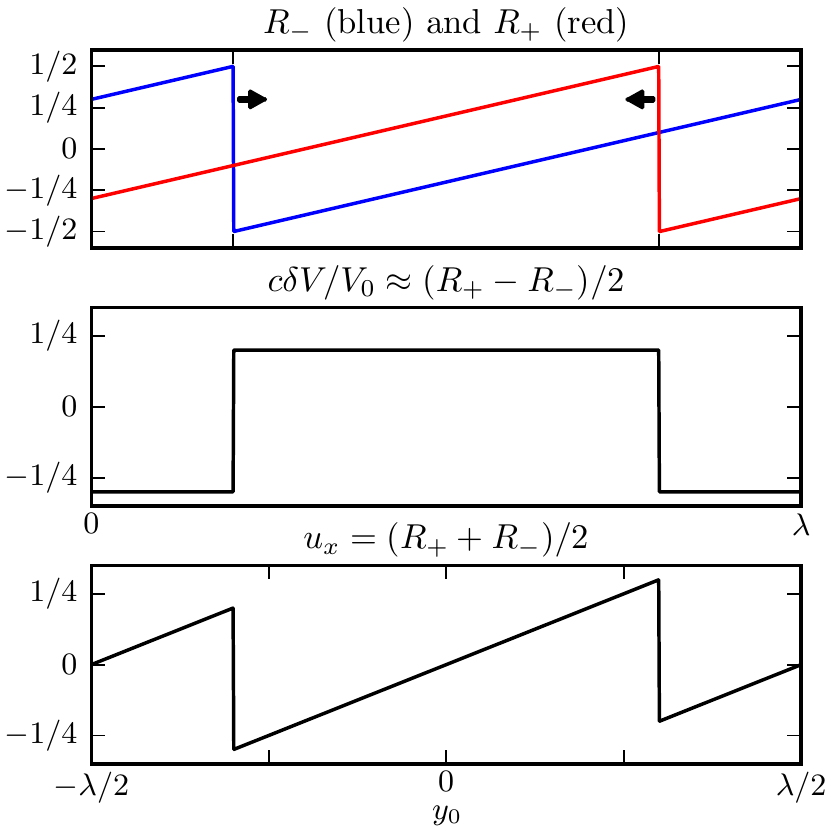}}
  \caption{Sketch of the superposition of two counter-propagating sawtooth
  waves. The arrows indicated the direction of propagation. Note that the
  waves travel towards a region of low pressure
  (since \mbox{$\delta{}p\sim1/\delta{}V$}).}
  \label{fig:sawtooth-profiles}
\end{figure}

In Figure~\ref{fig:sawtooth-profiles} we illustrate schematically the
superposition of two counter-propagating sawtooth waves, corresponding to the
weakly nonlinear asymptotic solutions derived in Section~\ref{sec:analytic}.
The arrows indicated the direction of propagation. Note that the forward and
backward propagating wave have the same slope. As we noted earlier, this is
because the latter is obtained from the former by simultaneously reversing the
sign of the wave amplitude and the spatial coordinate. Since the two sawtooth
waves have the same slope, subtracting one from the other yields a top hat
profile for the specific volume perturbation. Conversely, adding the two waves
results in a double sawtooth profile for the radial velocity perturbation.

\begin{figure}
  \centerline{\includegraphics{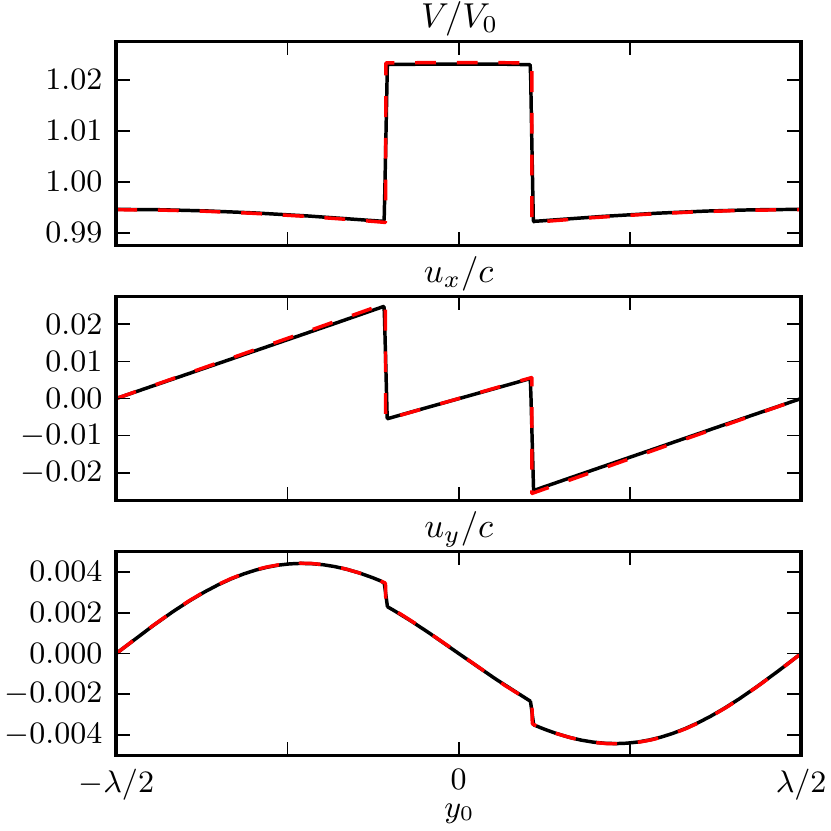}}
  \caption{Comparison of wave forms roughly three orbits after excitation
  (\mbox{$\Omega{}t\approx6\pi$}). Black: Numerical solution. Dotted red:
  Weakly nonlinear solution.}
  \label{fig:nonlinear-shape}
\end{figure}

We can gain additional confidence in the weakly nonlinear theory by testing it
directly against the numerical data. In Figure~\ref{fig:nonlinear-shape}, we
plot analytic solutions on top of numerical solutions at a time corresponding
to roughly three orbits after excitation. To achieve the best possible
agreement, we have incorporated in the analytic solution both the dispersive
correction to the sawtooth wave given by (\ref{eq:correction}) as well as the
stationary vortical wave that we used as initial condition.

As Figure~\ref{fig:nonlinear-shape} shows, the analytic solutions are
essentially indistinguishable from the numerical ones. We note that the only
free parameter that enters the analytic solutions is the location of the shock
front. The magnitude of the jump across the shock is in fact correctly
predicted by (\ref{eq:jump}). We also note that a superposition principle
evidently holds. This confirms what we have argued for in
Section~\ref{sec:wave-equation}, i.e.\ that weakly nonlinear,
counter-propagating sawtooth waves do not interact to a good approximation.

Close inspection of the wave profile of the specific volume perturbation
between shock fronts reveals that it is slightly curved. This curvature is a
dispersive effect that is correctly captured by the dispersive correction to
the leading order sawtooth solution. Unlike for $\delta{}V$, there is hardly
any curvature in the wave profile of the radial velocity $u_x$. This is
expected, as it turns out that the dispersive correction, which is quadratic
in the spatial coordinate, enters $R_{+}$ and $R_{-}$ with opposite sign, so
that it cancels out when their sum is taken to obtain $u_x$. We note that
these effects cannot be attributed to the vortical wave as its amplitude is
less than 1\% of the shock amplitude. In addition, the curvature in the
specific volume perturbation profile is everywhere concave. The vortical wave
on the other hand would give rise to to a sinusoidal variation that is concave
in one half of the domain and convex in the other.

Also illustrated in Figure~\ref{fig:nonlinear-shape} is the wave profile of
the azimuthal velocity $u_y$. The analytic solution shown consists of a
superposition of the vortical wave and the weakly nonlinear shock waves. The
wave form is mostly sinusoidal, indicating that the vortical wave dominates.
This is because in the case of $u_y$, the vortical wave decays as $t^{-1}$,
whereas the shock waves decay as $t^{-3}$, so that the former will always
dominate at late times.

For $\delta{}V$ and $u_x$, the vortical wave and the shock waves decay equally
fast (as $t^{-2}$). In general, one might therefore expect to see the vortical
wave at late times. Since this is not the case, we conclude that shortly after
the initial wave excitation, the shock wave should be much larger in amplitude
than the vortical wave. Inspection of Figure~\ref{fig:nonlinear-amplitudes}
shows that this is indeed the case.

\section{Discussion and application to MRI simulations}\label{sec:discussion}

We have considered the propagation of density waves in a shearing box. These
were assumed to be functions only of the shearing coordinate $y+q\Omega xt$
and time $t$. The waves were presumed to have been excited by vortensity
fluctuations produced by MRI turbulence as described by
\citet{HeinemannPapaloizou2009a,HeinemannPapaloizou2009b}. In this process it
is only the form of the vortensity fluctuation at the time when waves swing
from being leading to trailing that is significant. Just after the swing
inward and outward propagating waves are assumed, as found in simulations, to
be in the linear regime. Subsequently, as the waves propagate, their
wavelengths in the $x$ direction decreases, while the conservation of wave
momentum results in increased amplitude, causing them to enter the nonlinear
regime and the formation of shocks.

In Section~\ref{sec:analytic} of this paper we developed an analytic theory
that is applicable to the weakly nonlinear regime. We derived a modified
Burgers equation governing the dynamics of weakly nonlinear waves, which
contains terms describing nonlinear steepening, dispersion, and viscous
diffusion. We obtained nonlinear sawtooth solutions with weak shocks valid for
late times and estimated corrections resulting from dispersion.

In Section~\ref{sec:simulations} we presented numerical solutions of the
nonlinear equations without a small amplitude approximation. These led to
sawtooth waves with a profile that was in excellent agreement with that
obtained from the weakly nonlinear theory at late times, in particular with
regard to rate of decay of the shock velocity jump being ultimately
proportional to $t^{-2}$. The solutions constructed from the theory were
composed of a combination of forward and backward propagating waves, resulting
in a double sawtooth profile for the radial velocity $u_x$ and a top hat
profile for the specific volume perturbation $\delta{}V$.

\subsection{MRI simulations}

\begin{figure}
  \centerline{\includegraphics{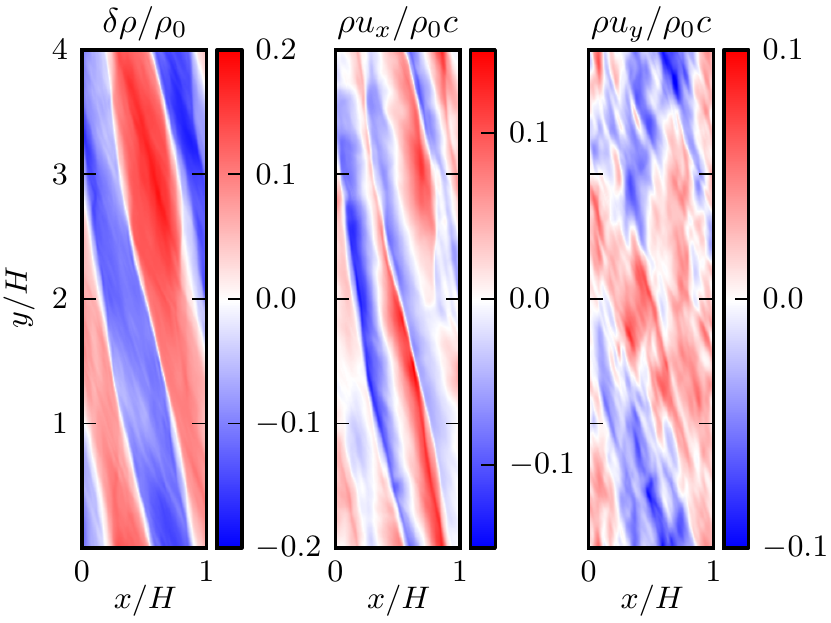}}
  \caption{Pseudo-color images of the vertically integrated mass and momentum
  densities taken from the simulation described in
  \citet{HeinemannPapaloizou2009b}.}
  \label{fig:mri-image}
\end{figure}

\begin{figure}
  \centerline{\includegraphics{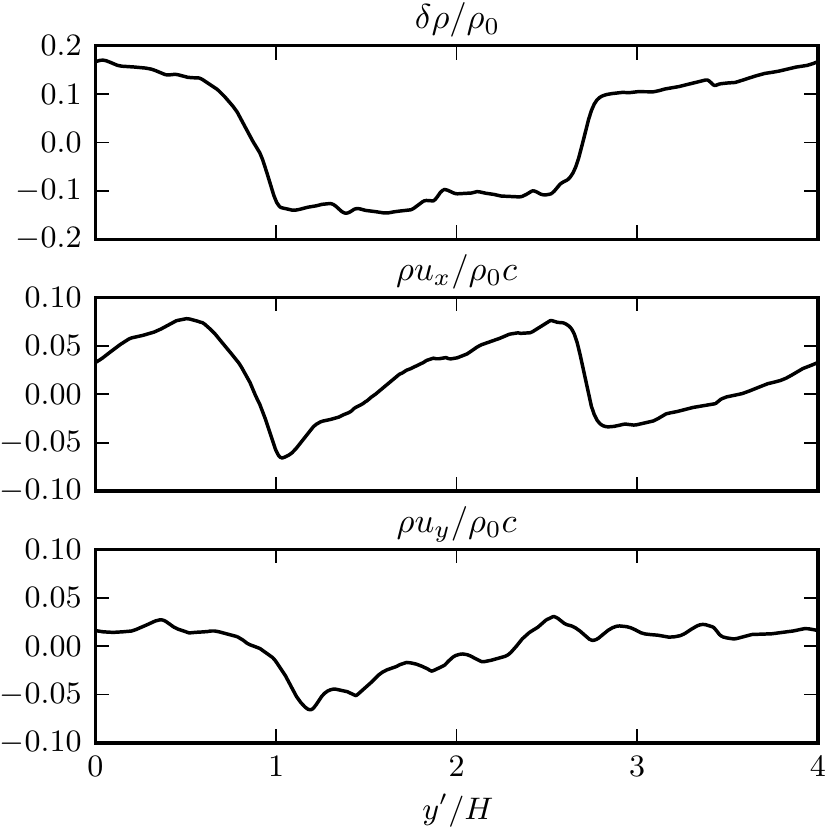}}
  \caption{Cut through Figure~\ref{fig:mri-image} along $y$ at a fixed
  $x\approx 0.15$ (no average over $x$).}
  \label{fig:mri-cut}
\end{figure}

An important issue is the extent to which the description of weakly nonlinear
waves we have provided above applies to the density waves excited in MRI
simulations. These waves are found to be ubiquitous in such simulations. As a
typical example of what is found in many realizations we show in
Figure~\ref{fig:mri-image} a snapshot of the (vertically integrated) mass and
momentum densities found in a simulation described in
\citet{HeinemannPapaloizou2009b}. From this figure it is apparent that the
wave structure is significantly more pronounced in $u_x$ and
\mbox{$\delta{}\rho/\rho_0\sim-\delta{}V/V_0$} than in the azimuthal velocity
$u_y$. This is expected from our discussion in
Section~\ref{sec:numerical-results}, which indicated that $u_y$ is dominated
by the stationary vortical wave. In addition, Figure~\ref{fig:mri-image}
indicates a top hat like profile in $\delta V$. To examine this more closely,
Figure~\ref{fig:mri-cut} shows plots of the fluid variables taken at a fixed
value of $x$. These may be compared to the corresponding plots in
Figure~\ref{fig:sawtooth-profiles}. It is seen that although there are
superposed fluctuations in the MRI simulation, there is good agreement between
the profiles: double sawtooth for $u_x$ and top hat for $\delta{}V$
(corresponding to an inverted top hat for the profile of $\delta\rho$). This
correspondence is less clear for $u_y$, but this is not unreasonable on account
of the expected dominance by vortical perturbations.

It should be noted that in an MRI simulation, spiral density waves propagate
through a field of turbulent fluctuations, and it might be expected that the
waves interact with these fluctuations. This could in general lead to wave
front distortion and energy loses resulting from random secondary density wave
generation through interaction of the primary wave with turbulent eddies
\citep[eg.][]{Lighthill1953,Howe1971b,Howe1971a,FfowcsWillamsHowe1973}. We
also remark that a spiral density wave is expected to undergo wave front
distortion as it propagates through a medium with local turbulent velocity
and/or local sound speed fluctuations \citep[eg.][]{HesselinkSturtevant1988}.
This may occur in a stationary medium with no generation of secondary waves
with different frequency.

We comment that a description of losses through wave-turbulence interactions
by an effective turbulent viscosity is rather problematic. Such a description
necessitates statistical averaging over many realizations of the system,
see e.g.\ \cite{BalbusPapaloizou1999}. Thus it most likely describes the
possible variation that different realizations can display rather than the
behavior of any one realization, where distortion of the wave profile rather
than diffusion may occur, an issue that leads to interpretational difficulties,
see \cite{FfowcsWillamsHowe1973} for an extensive discussion. These authors
also point out that an effective turbulent viscosity description fails to meet
the reasonable expectation that shock waves should be influenced mostly by
small scale turbulent motions.

In support of the above view, we comment that experimental results presented
by \citet{PlotkinGeorge1972} for wave profiles associated with sonic booms
propagating through a turbulent atmosphere indicate good agreement with non
turbulent theory apart from random perturbations to the wave profile and a
thickening of the shock front (that remains narrow on a macroscopic scale)
\citep[see also][]{GiddingsRusak2001}. Although the context differs, this view
would appear to be consistent with the results presented in
Figures~\ref{fig:mri-image} and \ref{fig:mri-cut}.

Nonetheless, it is likely that interaction with disorganized turbulence causes
the waves to decay more rapidly than predicted by our weakly nonlinear wave
theory. An inspection of the results in \citet{HeinemannPapaloizou2009b}
indeed indicates a decay rate faster than $t^{-2}$. However, this may be
affected by the particular choices of Reynolds and Prandtl numbers that are
made for numerical tractability. In this context it is important to note that
the decay rate produced in the simple weakly nonlinear wave theory enunciated
here gives, what we expect from the discussion above, to be a lower limit as
far as the simulations are concerned and it does not differ greatly from what
is actually observed. Thus the description of excited waves given by the
simple theory and the current simulations is unlikely to be greatly modified
when the transport coefficients are changed.

\section*{Acknowledgements}

Tobias Heinemann is supported by NSF grant AST–-0807432 and NASA grant
NNX08AH24G.

\bibliographystyle{mn2e}
\bibliography{nl-sd-wave}

\label{lastpage}

\end{document}